\documentclass[final,preprint,aps,showpacs,tightenlines,fleqn,
superscriptaddress, showkeys,nofootinbib]{revtex4}
\usepackage{graphicx}
\usepackage{bm}
\usepackage{dcolumn}
\begin{document}

\bibliographystyle{prsty}


\preprint{PNU-NURI-05/2004}
\preprint{PNU-NTG-04/2004}
\title{$\Delta S=0$ Effective weak chiral Lagrangian \\ from the instanton
vacuum}

\author{Hee-Jung Lee}
\email{hjlee@www.apctp.org}
\affiliation{Department of Physics and \\ Nuclear Physics \& Radiation Technology
Institute (NuRI), Pusan National University,
Busan 609-735, Republic of Korea}
\affiliation{Departament de Fisica Te\`orica, Universitat de Val\`encia
E-46100 Burjassot (Val\`encia), Spain}
\author{Chang Ho Hyun}
\email{hch@meson.skku.ac.kr}
\affiliation{School of Physics, Seoul National University,
Seoul 151-742, Republic of Korea}
\affiliation{Institute of Basic Science, Sungkyunkwan University,
Suwon 440-746, Republic of Korea}
\author{Chang-Hwan Lee}
\email{clee@pusan.ac.kr}
\affiliation{Department of Physics and \\ Nuclear Physics \& Radiation Technology
Institute (NuRI), Pusan National University,
Busan 609-735, Republic of Korea}
\author{Hyun-Chul Kim}
\email{hchkim@pusan.ac.kr}
\affiliation{Department of Physics and \\ Nuclear Physics \& Radiation Technology
Institute (NuRI), Pusan National University,
Busan 609-735, Republic of Korea}
\date{August 9, 2005}

\begin{abstract}
We investigate the $\Delta S=0$ effective chiral Lagrangian from the
instanton vacuum.  Based on the $\Delta S=0$ effective weak
Hamiltonian from the operator product expansion and renormalization
group equations, we derive the strangeness-conserving effective weak
chiral Lagrangian from the instanton vacuum to order ${\cal O}(p^2)$
and the next-to-leading order in the $1/N_c$ expansion at the quark
level.  We find that the quark condensate and a dynamical term
which arise from the QCD and electroweak penguin operators appear in
the next-to-leading order in the $1/N_c$ expansion for the $\Delta S
=0$ effective weak chiral Lagrangian, while they are in the leading
order terms in the $\Delta S = 1$ case.  Three different types of the
form factors are employed and we find that the dependence on the
different choices of the form factor is rather insensitive.  The
low-energy constants of the Gasser-Leutwyler type are determined and
discussed in the chiral limit.
\end{abstract}
\maketitle
\setcounter{footnote}{0}

\section{Introduction}
A great deal of attention has been paid to parity violation
(PV) in the electroweak standard model (SM) well over decades in the
context of high-precision tests for the SM~\cite{Bouchiat:kt}.  A
recent series of parity-violating
experiments in atomic physics measured the weak charge of the
SM~\cite{Wood:zq,Bennett:1999pd,Meekhofetal,Vetter:vf}.  Its
discrepancy with the SM implies a possibility of new physics, for
example, a possible existence of the $Z'$ boson in addition to the
$Z^0$ boson~\cite{Casalbuoni:1999yy,Rosner:1999cy,Erler:1999nx}.
Recently, strangeness-conserving ($\Delta S=0$) weak processes
have paved the way of probing subtle properties of the nucleon such as
the strangeness in the nucleon: The strange vector form
factors were recently extracted by measuring the asymmetries of PV
$\vec{e}p$ parity-violating scattering~\cite{epscat}.  Hadronic and
nuclear PV processes, however, are far from clear understanding due to
the screening of the strong interaction.

A simple framework to describe hadronic and nuclear PV processes is
one-boson exchange (OBE) such as $\pi$-, $\rho$-, and
$\omega$-changes ~\cite{Desplanques:1979hn,Miller:1982ij,Dubovik:pj}
{\em \`{a} la} the strong nucleon-nucleon potential through
OBE~\cite{Nagels:fb,Machleidt:hj}.  The main ingredients of the PV OBE
model are the weak meson-nucleon coupling constants such as
$h_\pi$, $h_\rho$, and $h_\omega$, which can be extracted from PV
observables in various hadronic and nuclear reactions like $\vec{p}p$
elastic scattering~\cite{ppscat},
$\vec{n}p\rightarrow d\gamma$~\cite{Cavaignac:uk,Snow:az}, and
$^{18}{\rm F}^*\rightarrow \hspace{-8pt}\phantom{0}^{18}{\rm
  F}$~\cite{f18}.  In particular, the weak
pion-nucleon coupling constant $h^1_\pi$ is one of the most
important quantities dominant in PV weak hadronic processes
at low-energy
regions~\cite{Desplanques:1979hn,Adelberger:1985ik,Kaplan:1992vj,
Chen:2000km}.  However, disagreement in determining the
$h^1_\pi$ still exists~\cite{Wilburn:1998xq} theoretically as
well as experimentally.

A recent series of
works~\cite{Kaiser:1988bt,Kaiser:1989fd,Meissner:1998pu} studied the
$h^1_\pi$ within the Skyrme model, based on the
effective current-current interaction which can be identified as a
factorization scheme.  However, it is natural to describe the $\Delta
S=0$ PV processes based on the effective weak Hamiltonian evolved from
a scale of 80 GeV down to around 1
GeV~\cite{Altarelli:1974ni,Galic:ke,Donoghue:xk,Desplanques:1979hn,
Dai:bx}.  Furthermore, it is well known that the nonleptonic weak
processes defy any explanation from the factorization, or the strict
large-$N_c$ limit.  The octet enhancement in $K\rightarrow \pi\pi$
decays is partially explained by gluon penguin diagrams, which
indicates that the strong interaction plays an essential role in
describing the nonleptonic or hadronic decay processes.  Thus,
in the present work, we shall derive the $\Delta S=0$ effective weak chiral
Lagrangian (EW$\chi$L) incorporating the effective weak
Hamiltonian~\cite{Desplanques:1979hn}, based on the nonlocal chiral
quark model from the instanton vacuum~\cite{DP}, which will provide a good
theoretical framework in studying the weak coupling
constants~\cite{Leeetal}.  We shall consider the $\Delta S=0$ EW$\chi$L
to order ${\cal   O}(p^2)$ in the chiral limit and to the
next-to-leading order (NLO) in the $1/N_c$ expansion, keeping in mind
that the present results of the NLO in $N_c$ corrections are just a
part of the whole $1/N_c$ NLO contributions.

The nonlocal chiral quark model induced from the instanton vacuum has
several virtues: It was shown that this momentum dependence gives the
correct end-point behavior of the quark virtuality for the pion wave
function~\cite{Petrov:1998kg,Praszalowicz:2001wy}. Similarly, recent
investigations on the effective weak chiral
Lagrangian~\cite{Franz:1999wr,Franz:1999yz,Franz:1999ik} indicate that
the momentum-dependent quark mass plays a significant role in
enhancing the $\Delta T=1/2$ channel.  Furthermore,
nonlocality of the quark introduces a unique feature to the low-energy
constants (LEC)~\cite{Choi:2003cz}, compared to other models.

The paper is organized as follows : In Section II, we show how to
incorporate the $\Delta S=0$ effective weak Hamiltonian. In Section
III, we discuss the present results of the low-energy constants.  In
particular, the behavior of the LEC is studied with
respect to the momentum-dependent quark mass.  In the last section, we
summarize the present work and draw conclusions.

\section{$\Delta S=0$ effective weak chiral Lagrangian}
In this section, we will show how to incorporate the $\Delta S=0$
effective weak Hamiltonian into the effective chiral action from the
instanton vacuum.  The detailed description can be found in
Ref.~\cite{Franz:1999ik}.  We employ the $\Delta S=0$ effective PV
weak Hamiltonian derived in Ref.~\cite{Desplanques:1979hn}.  The
Hamiltonian reads
\begin{eqnarray}
{\cal H}_{W}^{\Delta S=0} &=&
\frac{G_F}{\sqrt{2}}\cos\theta_c\sin\theta_c \bigg[
\sum_{i,j=1}^2(\alpha_{ij}{\cal O}(A_i^\dagger, A_j)
+\beta_{ij}{\cal O}(A_i^\dagger t_A, A_j t_A)+{\rm h.c.}))\nonumber\\
&&+\sum_{i,j=1}^2(\gamma_{ij}{\cal O}(B_i^\dagger, B_j)
+\rho_{ij}{\cal O}(B_i^\dagger t_A, B_j t_A)) \bigg],
\label{Heff}
\end{eqnarray}
where the operator ${\cal O}(M,N)$ is defined as
${\cal O}(M,N) \equiv -{\psi}^\dagger\gamma_\mu\gamma_5 M \psi
{\psi}^\dagger\gamma^\mu N\psi$ in Euclidean space, and $t_A$ denotes
the generator of the color $SU(3)$ group, normalized as ${\rm tr}\,
t_At_B=2\delta_{AB}$.  The definitions of the matrices $A_i$ and
$B_i$, and the coefficients $\alpha$, $\beta$, $\gamma$ and $\rho$ are
given in \cite{Desplanques:1979hn}.  These coefficients are the
functions of the scale-dependent Wilson coefficient $K(\mu)$ defined
as
\begin{equation}
K(\mu) \equiv
\bigg(1+\frac{g^2(\mu^2)}{16\pi^2}b\ln\frac{M_W^2}{\mu^2}\bigg),
\end{equation}
where $g$ is the strong coupling constant, $\mu$ is the
renormalization point and specifies the mass scale, $b=11-2 N_f/3$, and
$M_W$ is the mass of the $W$ boson.  $K$ encodes the effect of the strong
interaction from the perturbative gluon exchanges.  Numerical values
of the coefficients relevant to our discussion with various $K$ values
are listed in Tab.~\ref{Coeff}.
\begin{table}
\begin{center}
\begin{tabular}{|c|c|c|c|}
\hline
     &    $K=1$   &   $K=4$    & $K=7$ \\ \hline\hline
$\alpha_{11}$ & $\cot\theta_c$ & $1.126\cot\theta_c$ & $1.266\cot\theta_c$ \\
\hline
$\alpha_{22}$ & $\tan\theta_c$ & $1.126\tan\theta_c$ & $1.266\tan\theta_c$ \\
\hline
$\beta_{11}$  & $0$ & $-0.307\cot\theta_c$ & $-0.479\cot\theta_c$ \\
\hline
$\beta_{22}$  & $0$ & $-0.307\tan\theta_c$ & $-0.479\tan\theta_c$ \\
\hline
$\gamma_{11}$ & $-0.002(1-\frac{2}{3}\sin^2\theta_w)\csc2\theta_c$
& $0.077(1-\frac{2}{3}\sin^2\theta_w)\csc2\theta_c$
& $0.163(1-\frac{2}{3}\sin^2\theta_w)\csc2\theta_c$ \\
\hline
$\gamma_{12}$ & $0.007\sin^2\theta_w\csc2\theta_c$ &
     $0.001\sin^2\theta_w\csc2\theta_c$
& $0.006\sin^2\theta_w\csc2\theta_c$ \\
\hline
$\gamma_{21}$ & $-0.671\sin^2\theta_w\csc2\theta_c$ &
     $-0.772\sin^2\theta_w\csc2\theta_c$
& $-0.898\sin^2\theta_w\csc2\theta_c$ \\
\hline
$\gamma_{22}$ & $(1-2\sin^2\theta_w)\csc2\theta_c$
& $1.101(1-2\sin^2\theta_w)\csc2\theta_c$ &
     $1.236(1-2\sin^2\theta_w)\csc2\theta_c$ \\
\hline
$\rho_{11}$ & $0$ & $-0.190(1-\frac{2}{3}\sin^2\theta_w)\csc2\theta_c$
& $-0.296(1-\frac{2}{3}\sin^2\theta_w)\csc2\theta_c$ \\
\hline
$\rho_{12}$ & $0.003\sin^2\theta_w\csc2\theta_c$ &
     $0.260\sin^2\theta_w\csc2\theta_c$
& $0.453\sin^2\theta_w\csc2\theta_c$ \\
\hline
$\rho_{21}$ & $-0.001\sin^2\theta_w\csc2\theta_c$ &
     $0.002\sin^2\theta_w\csc2\theta_c$
& $-0.032\sin^2\theta_w\csc2\theta_c$ \\
\hline
$\rho_{22}$ & $0$ & $-0.307(1-2\sin^2\theta_w)\csc2\theta_c$
& $-0.479(1-2\sin^2\theta_w)\csc2\theta_c$ \\
\hline
\end{tabular}
\caption{Strong enhancements with selected values of $K$.
$\theta_w$ and $\theta_c$ are the Weinberg and the Cabbibo
angles, respectively.
\label{Coeff}}
\end{center}
\end{table}
We denote the four-quark operators generically by
\begin{equation}
{\cal Q}^i(x)=-{\psi}^\dagger(x)\Gamma_1^i\psi(x)
{\psi}^\dagger(x)\Gamma_2^i\psi(x)\ ,
\end{equation}
where $i(=1\cdots 12)$ labels each four-quark operator in the effective
weak Hamiltonian and $\Gamma_{1(2)}^i$ consist of the $\gamma$
and the flavor matrices.  Thus, the effective weak Hamiltonian can be
rewritten as follows:
\begin{equation}
{\cal H}^{\Delta S=0}_{W}
=\sum_{i=1}^{12}{\cal C}_i {\cal Q}^i(x),
\end{equation}
where ${\cal C}_i$ denotes $\alpha$, $\beta$, $\gamma$ and
$\rho$ in Eq.~(\ref{Heff}).

The $\Delta S=0$ effective PV weak Hamiltonian can be incorporated into the
nonlocal chiral quark model as follows:
\begin{equation}
\exp(-S_{\rm eff}^{\Delta S=0})=
\int {\cal D}\psi {\cal D}\psi^\dagger\exp\bigg(\int d^4x(\psi^\dagger
D \psi-{\cal H}_{W}^{\Delta S=0})\bigg),
\label{eq:effac}
\end{equation}
where the $D$ denotes the nonlocal Dirac operator defined by
\begin{equation}
D(-i\partial) \equiv
i\gamma_\mu\partial_\mu+i\sqrt{M(-i\partial)}U^{\gamma_5}(x)\sqrt{M(-i\partial)}.
\end{equation}
Since the Fermi constant $G_F$ is very small, we can expand the
exponent in Eq.~(\ref{eq:effac}) in powers of $G_F$ and keep the lowest order
only.  Thus, the $\Delta S=0$ EW$\chi$L can be
derived as
\begin{eqnarray}
{\cal L}_{\rm eff}^{\Delta S=0}
&=&\int {\cal D}\psi {\cal D}\psi^\dagger{\cal H}_{W}^{\Delta S=0}
\exp\int d^4z\ \psi^\dagger(z) D\psi(z).
\end{eqnarray}
The vacuum expectation value (VEV) of the four-fermion operators in
the effective weak Hamiltonian can be calculated as
\begin{eqnarray}
\langle {\cal Q}^i(x)\rangle
&=&-\frac{1}{\cal Z} \int d^4y\delta^4(x-y) \frac{\delta}{\delta
  J_1^i(x)}\frac{\delta}{\delta J_2^i(y)}\int {\cal D}\psi {\cal
  D}\psi^\dagger
\nonumber\\
&&\times\exp\int d^4z\ \psi^\dagger(z)(D+J_1^i(z)\Gamma_1^i
+J_2^i(z)\Gamma_2^i)\psi(z)
\bigg|_{J_1=J_2=0}\nonumber\\
&=& {\rm tr}_{c,\gamma,f}\bigg[\langle x|D^{-1}\Gamma^{(i)}_1|x\rangle
\langle x|D^{-1}\Gamma^{(i)}_2|x\rangle\bigg]
\nonumber \\
&-&{\rm tr}_{c,\gamma,f}\bigg[\langle x|D^{-1}\Gamma^{(i)}_1|x\rangle\bigg]
{\rm tr}_{c,\gamma,f}\bigg[\langle x|D^{-1}\Gamma^{(i)}_2|x\rangle\bigg]\ ,
\label{Qaverage}
\end{eqnarray}
where ${\rm tr}_{c\gamma f}$ means the trace over color, spin, and
flavor space, respectively.  The last two lines in
Eq.~(\ref{Qaverage}) correspond to the unfactorized and factorized
quark loops, respectively.
$\langle x|(D)^{-1}\Gamma^{(i)}_{1, 2}|x\rangle$ can be easily
calculated as
\begin{equation}
\langle x|D^{-1}\Gamma^{(i)}_l|x\rangle
=\int \frac{d^4k}{(2\pi)^4}\
\frac{1}{D^{\dagger}(\partial+ik)D(\partial+ik)}
D^{\dagger}(\partial+ik)\Gamma^{(i)}_l \ .
\label{eq:dg}
\end{equation}
To order ${\cal O}(k^2)$ can be expanded the denominator of
Eq.(\ref{eq:dg}) as follows:
\begin{eqnarray}
&&D^\dagger(\partial+ik)D(\partial+ik)\nonumber\\
&=&-\partial^2+k^2-2ik\cdot\partial+M^2-M(\gamma_\mu\partial_\mu U^{\gamma_5})
-2iM\tilde{M}^\prime k_\mu[2\partial_\mu+U^{-\gamma_5}(\partial_\mu U^{\gamma_5})]
\nonumber\\
&&-M\tilde{M}^\prime [2\partial^2
+U^{-\gamma_5}(\partial^2 U^{\gamma_5})
+2U^{-\gamma_5}(\partial_\mu U^{\gamma_5})\partial_\mu]
\nonumber\\
&&-2M\tilde{M}^{\prime\prime}k_\mu k_\nu[2\partial_\mu\partial_\nu
+U^{-\gamma_5}(\partial_\mu\partial_\nu U^{\gamma_5})+
2U^{-\gamma_5}(\partial_\mu U^{\gamma_5})\partial_\nu]
\nonumber\\
&&-2\tilde{M}^{\prime 2}k_\mu k_\nu
[(\partial_\mu U^{-\gamma_5})(\partial_\nu U^{\gamma_5})
+U^{-\gamma_5}(\partial_\mu\partial_\nu U^{\gamma_5})
+2U^{-\gamma_5}(\partial_\mu U^{\gamma_5})\partial_\nu
+ 2\partial_\mu\partial_\nu]
\nonumber\\
&&+i\tilde{M}^{\prime}k_\mu[(\partial_\mu\, \gamma \cdot \partial U^{\gamma_5})
+2(\gamma \cdot \partial U^{\gamma_5})\partial_\mu]+{\cal O}(\partial^3)
\label{DD}
\end{eqnarray}
The numerator reads
\begin{equation}
D^\dagger(\partial+ik)=i\gamma_\mu(\partial_\mu+ik_\mu)-iB,
\end{equation}
where
\begin{eqnarray}
B&=&M(k)U^{-\gamma_5}-i\tilde{M}^{\prime}k\cdot(\partial U^{-\gamma_5})
-\bigg({M}^{\prime\prime}-\frac{\tilde{M}^{\prime 2}}{2M}\bigg)
k_\alpha k_\beta(\partial_\alpha\partial_\beta U^{-\gamma_5})
-\frac{\tilde{M}^\prime}{2}.
\label{B}
\end{eqnarray}
Therefore, we have
\begin{eqnarray}
\langle x|D^{-1}\Gamma^{(i)}_l|x\rangle
&=&\int \frac{d^4k}{(2\pi)^4}\ \frac{1}{k^2+M^2(k)-A}
(i\gamma_\mu k_\mu-B)(i\Gamma^{(i)}_l)
\nonumber\\
&=&\sum_{n=0}^{\infty}\int \frac{d^4k}{(2\pi)^4}\ \frac{1}{k^2+M^2(k)}
\bigg(\frac{1}{k^2+M^2(k)}A\bigg)^n
(i\gamma_\mu k_\mu-B)(i\Gamma^{(i)}_l),
\label{eq:expand1}
\end{eqnarray}
where the form of $A$ can be extracted from Eq.~(\ref{DD}).
The expansion of Eq.~(\ref{eq:expand1}) yields the terms to order
${\cal O}(\partial^2)$:
\begin{eqnarray}
&&\langle x|D^{-1}\Gamma^{(i)}_l|x\rangle \cr
&=&\bigg({\cal I}_{1} U^{-\gamma_5}
+{\cal I}_{2}U^{-\gamma_5}(\partial_\alpha U^{\gamma_5})\gamma_\alpha
+{\cal I}_{3} (\partial^2U^{-\gamma_5})
+{\cal I}_{4} U^{-\gamma_5}(\partial_\alpha U^{\gamma_5})(\partial_\beta
U^{\gamma_5^\dagger})\gamma_\alpha\gamma_\beta\bigg)i\Gamma_l^{(i)}\,
\label{DGamma}
\end{eqnarray}
with the coefficients
\begin{eqnarray}
{\cal I}_{1}&=&-\int \frac{d^4 k}{(2\pi)^4}\ \frac{M(k)}{k^2+M^2(k)}
=\frac{\left<\overline{\psi}{\psi}\right>_{\rm M}}{4N_c},
\label{eq:coefficient1} \\
{\cal I}_{2}&=&\int \frac{d^4 k}{(2\pi)^4}\ \frac{M^2(k)
-\frac{k^2}{2}M(k)\tilde{M}^\prime}{(k^2+M^2(k))^2},
\label{eq:coefficient2}\\
{\cal I}_{3}&=&\int \frac{d^4 k}{(2\pi)^4}\
\bigg[\frac{\frac{1}{4}\tilde{M}^{\prime\prime}k^2
+\frac{1}{2}\tilde{M}^\prime-\frac{\tilde{M}^{\prime2}}{8M}k^2}{k^2+M^2(k)}
\cr
&&-\frac{M+M^2\tilde{M}^\prime+\frac{k^2}{2}M^2\tilde{M}^{\prime\prime}
+\frac{1}{2}k^2M\tilde{M}^{\prime2}
+\frac{k^2}{4}\tilde{M}^\prime}{(k^2+M^2(k))^2} \cr
&&
+k^2\frac{\frac{1}{2}M+2M^2\tilde{M}^\prime
+M^3\tilde{M}^{\prime2}}{(k^2+M^2(k))^3}\bigg],\label{eq:coefficient3} \\
{\cal I}_{4}&=&\int \frac{d^4 k}{(2\pi)^4}\
\frac{-M^3+k^2M^2\tilde{M}^\prime}{(k^2+M^2(k))^3}\ .
\label{eq:coefficient4}
\end{eqnarray}
Substituting Eq.~(\ref{DGamma}) into Eq.~(\ref{Qaverage}), taking trace
over color and spin spaces, and summing all four-fermion operators,
we arrive at the $\Delta S=0$ EW$\chi$L to order
${\cal O}(\partial^2)$ in terms of the Goldstone boson fields with the
LEC determined:
\begin{eqnarray}
{\cal L}_{\rm eff}^{\Delta S=0}
&=& {\cal N}_{1} \bigg( \left<(R_\mu-L_\mu)\lambda_1\right>
\left<(R^\mu+L^\mu)\lambda_1\right>
+\left<(R_\mu-L_\mu)\lambda_2\right>
\left<(R^\mu+L^\mu)\lambda_2\right>\bigg)
\nonumber\\
&+& {\cal N}_{2} \bigg(\left<(R_\mu-L_\mu)\lambda_4\right>
\left<(R^\mu+L^\mu)\lambda_4\right>
+\left<(R_\mu-L_\mu)\lambda_5\right>
\left<(R^\mu+L^\mu)\lambda_5\right>\bigg)
\nonumber\\
&+& {\cal N}_3
\left<R_\mu-L_\mu\right>\left<R^\mu+L^\mu\right>
\nonumber\\
&+& {\cal N}_4 \left<R_\mu-L_\mu\right>
\left<(R^\mu+L^\mu)(-\frac{I}{3}+\lambda_3+\frac{1}{\sqrt{3}}\lambda_8)\right>
\nonumber\\
&+&{\cal N}_{5}\left<(R_\mu-L_\mu)(-\frac{I}{3}+\lambda_3+\frac{1}{\sqrt{3}}\lambda_8)\right>
\left<R^\mu+L^\mu\right>
\nonumber\\
&+& {\cal N}_{6} \left<(R_\mu-L_\mu)
(-\frac{I}{3}+\lambda_3+\frac{1}{\sqrt{3}}\lambda_8)\right>
\left<(R^\mu+L^\mu)
(-\frac{I}{3}+\lambda_3+\frac{1}{\sqrt{3}}\lambda_8)\right>
\nonumber\\
&+& {\cal N}_{7} \left<R_\mu\lambda_1 R^\mu\lambda_1
-L_\mu\lambda_1 L^\mu\lambda_1
+R_\mu\lambda_2 R^\mu\lambda_2-L_\mu\lambda_2
L^\mu\lambda_2\right>
\nonumber\\
&+& {\cal N}_{8} \left<R_\mu\lambda_4 R^\mu\lambda_4
-L_\mu\lambda_4 L^\mu\lambda_4
+R_\mu\lambda_5 R^\mu\lambda_5-L_\mu\lambda_5 L^\mu\lambda_5\right>
\nonumber\\
&+& {\cal N}_9 \left<(R_\mu R^\mu-L_\mu
L^\mu)(\lambda_3+\frac{1}{\sqrt{3}}\lambda_8)\right>
\nonumber\\
&+& {\cal N}_{10} \left<R_\mu(\lambda_3+\frac{1}{\sqrt{3}}\lambda_8)
R^\mu(\lambda_3+\frac{1}{\sqrt{3}}\lambda_8)
-L_\mu(\lambda_3+\frac{1}{\sqrt{3}}\lambda_8)
L^\mu(\lambda_3+\frac{1}{\sqrt{3}}\lambda_8)\right>
\label{Leff}
\end{eqnarray}
where $R_\mu \equiv iU\partial_\mu U^\dagger$,
$\left<\cdots\right>$ represents again the trace over flavor
space, and ${\cal N}_i$ denote the LEC expressed as follows:
\begin{eqnarray}
{\cal N}_1 &=& 2 N_c^2 {\cal I}_2^2\tilde{\alpha}_{11}, \;\;\;\;
{\cal N}_2  =  2 N_c^2 {\cal I}_2^2\tilde{\alpha}_{22},\;\;\;\;
{\cal N}_3  =  4 N_c^2 {\cal I}_2^2\tilde{\gamma}_{11},\cr
{\cal N}_4  &=&  4 N_c^2 {\cal I}_2^2\tilde{\gamma}_{12},\;\;\;\;
{\cal N}_{5}= 4 N_c^2 {\cal I}_2^2 \tilde{\gamma}_{21},\;\;\;\;
{\cal N}_{6}=  4 N_c^2 {\cal I}_2^2\tilde{\gamma}_{22},\cr
{\cal N}_7  &=&  2 N_c   {\cal I}_2^2(\tilde{\alpha}_{11}
+2\tilde{\beta}_{11}),\;\;\;\;
{\cal N}_8  =  2 N_c   {\cal I}_2^2(\tilde{\alpha}_{22}
+2\tilde{\beta}_{22}),\cr
{\cal N}_9 &=& 4 N_c\bigg[ 4{\cal I}_1{\cal I}_3
(\tilde{\gamma}_{12}+\tilde{\gamma}_{21}+2\tilde{\rho}_{12}+2\tilde{\rho}_{21})
+{\cal I}_2^2
(\tilde{\gamma}_{12}-\tilde{\gamma}_{21}+2\tilde{\rho}_{12}-2\tilde{\rho}_{21})
\bigg],\cr
{\cal N}_{10}  &=&   4 N_c {\cal I}_2^2(\tilde{\gamma}_{22}
+2\tilde{\rho}_{22}).
\label{LECs}
\end{eqnarray}
Here, $\tilde{{\cal C}}_{ij}$ stand for
$\frac{G_F}{\sqrt{2}}\sin\theta_c\cos\theta_c{\cal C}_{ij}$
generically, where ${\cal C}_{ij} = \alpha,\, \beta,\, \gamma,\, \rho$
in Eq.~(\ref{Heff}).  Note that the $\alpha_{ij}$ and $\gamma_{ij}$ enter in the
leading order (LO) Lagrangian, while the $\beta_{ij}$ and $\rho_{ij}$
appear only in the subleading order in $N_c$.  The numerical
evaluation of the LEC will be discussed in the next Section.
\section{Results and Discussions}
The large $N_c$ expansion in the context of nonleptonic decays have
been discussed already extensively~\cite{Fukugita,Tadicetal,Chivukula,
Bardeenetal,Goity,PichRafael2}.  While the large $N_c$ argument
works very well in the strong interaction, it does not seem
to describe the nonleptonic weak interactions in the leading order
(LO) of the large $N_c$ expansion.  The strict $1/N_c$
expansion is identical to a naive factorization: There is no mixing
in the operators and it leaves only the original four-quark operator which
contains the product of two conserved currents.  Thus, one has to
consider the NLO in the $1/N_c$ expansion.
However, if the NLO contribution is large, a problem of its
convergence would arise. Moreover, there are various sources of the
NLO corrections in the large $N_c$ expansion such as mesonic loop
contributions.  We are not in a position to take into account all
possible NLO corrections in this work.  Thus, we will restrict our
scheme in the following: First, we will treat the Wilson coefficients in
a more practical way, i.e. we will not consider the $N_c$ behavior of
the Wilson coefficients.  Second, we consider the NLO in the $1/N_c$
expansion at the quark level.  It does not mean that these corrections
are more important or favorable, compared to other $1/N_c$ corrections
such as mesonic loop corrections.  We only intend in the present work
to compare the LO contribution with the NLO corrections at the quark
level.  By doing that, we will see that the structure of the $\Delta
S=0$ EW$\chi$L is rather different from the $\Delta
S=1$ EW$\chi$L.

We first consider the $\Delta S=0$ EW$\chi$L in the
LO of $N_c$, and investigate its behavior with respect to the form
factors and the Wilson coefficients.
In the large $N_c$ limit, the EW$\chi$L becomes
\begin{eqnarray}
{\cal L}_{\rm eff}^{\Delta S=0,N^2_c} &=& \frac{16{\cal
I}_2^2N_c^2}{f_\pi^4}\left[2\bigg( \tilde{\alpha}_{11}\sum_{i=1}^2V_\mu^i
A^{i \mu} +\tilde{\alpha}_{22}\sum_{i=4}^5V_\mu^i A^{i \mu}\bigg)
+9\tilde{\gamma}_{11}A_\mu^0 V^{0 \mu} \right.
\nonumber\\
&&+3\tilde{\gamma}_{12}
\bigg(-V_\mu^0+2V_\mu^3+\frac{2}{\sqrt{3}}V_\mu^8\bigg)
A^{0 \mu}
+3\tilde{\gamma}_{21}
\bigg(-A_\mu^0+2A_\mu^3+\frac{2}{\sqrt{3}}A_\mu^8\bigg)
V^{0 \mu}
\nonumber\\
&&\left.+\tilde{\gamma}_{22}
\bigg(-V_\mu^0+2V_\mu^3+\frac{2}{\sqrt{3}}V_\mu^8\bigg)
\bigg(-A^{0 \mu}+2A^{3 \mu}+\frac{2}{\sqrt{3}}A^{8 \mu}\bigg)
\right],
\label{L-leading1}
\end{eqnarray}
where $V^a_\mu$ and $A^a_\mu$ are the vector and axial-vector
currents, respectively, defined as
\begin{equation}
V_\mu^a =\frac{f_\pi^2}{2} \left\langle T^a(R_\mu + L_\mu)\right\rangle,\ \ \
A_\mu^a =\frac{f_\pi^2}{2} \left\langle T^a(R_\mu - L_\mu)\right\rangle.
\end{equation}
$T^a$ is the generator of the $U_f(3)$,
$T^a=(\frac{1}{3},\frac{\lambda^1}{2},\cdots,\frac{\lambda^8}{2}).$

The EW$\chi$L given in Eq.~(\ref{L-leading1})
has one caveat: In the large $N_c$ limit the four-quark operators turn
out to be products of two conserved currents, i.e. the vector and the
axial-vector currents.  However, the presence of the nonlocal
interaction between quarks and Goldstone bosons, which arises from the
momentum-dependent quark mass, breaks the gauge invariance, so that
the currents are not conserved.  Ref.~\cite{Franz:1999ik}
discussed a method of how to avoid this problem.  The conserved
currents in Euclidean space with the nonlocal interactions can be
derived by gauging the partition function.  The pion decay constant
$f_\pi^2$ can be successfully reproduced by using
the modified axial-vector current in the following matrix elements:
\begin{equation}
\left<0\left|A_\mu^a (x)\right| \pi^b(p)\right> = i f_\pi p_\mu
e^{ip\cdot x} \delta^{ab},
\end{equation}
which indicates that the Takahashi-Ward identity of PCAC is well
satisfied with the modified conserved axial-vector current.  If we use
the usual currents such as $A_\mu^a =
\bar{\psi}\gamma_\mu\gamma_5\lambda^a\psi$, we would end up with the
Pagels-Stokar expression for $f_\pi^2$:
\begin{equation}
f_\pi^2 ({\rm PS}) = 4N_c\int \frac{d^4 k}{(2\pi)^4} \frac{M^2-\frac14
  MM' k}{(k^2+M^2)^2}.
\label{eq:ps}
\end{equation}
Thus, one has to consider the modified conserved currents in
Eq.~(\ref{L-leading1}).  However, if we use the $f_\pi^2$(PS)
for the normalization of the effective chiral Lagrangian for
convenience, we need not introduce them in Eq.~(\ref{L-leading1}),
since we derive the same results as we use the modified conserved
currents.  Thus, the prefactor $16{\cal I}_2^2N_c^2$ in
Eq.~(\ref{L-leading1}) turns out to be $f_\pi^4$.

Moreover, in the strict large $N_c$ limit, the original Cabbibo and
Weinberg-Salam Lagrangians need not any renormalization,
i.e. the terms with $\alpha_{11}$, $\alpha_{22}$, $\gamma_{21}$,
and $\gamma_{22}$ survive.  Thus, the $\Delta S=0$ EW$\chi$L 
in the large $N_c$ limit becomes as follows:
\begin{eqnarray}
{\cal L}_{\rm eff}^{\Delta S=0,N_c^2}
&=&\sqrt{2}G_F\bigg\{\cos^2\theta_c\sum_{i=1}^2V_\mu^i A^{i
  \mu} +\sin^2\theta_c\sum_{i=4}^5V_\mu^i A^{i \mu}
\nonumber\\
&&-\bigg(\cos2\theta_w(V_\mu^3+\frac{1}{\sqrt{3}}V_\mu^8)
-\frac{1}{2}V_\mu^0\bigg)
\bigg(\frac{1}{2}A^{0 \mu}-A^{3 \mu}
-\frac{1}{\sqrt{3}}A^{8 \mu}\bigg)\bigg\}.
\label{LK=1}
\end{eqnarray}
If we take the limit $\theta_c\rightarrow 0$, Eq.~(\ref{LK=1}) becomes
identical to that in Ref.~\cite{Meissner:1998pu}.  However, if we take a
more practical point of view about the large $N_c$ behavior of the
anomalous dimensions~\cite{Franz:1999ik}, we get the $\Delta S=0$
EW$\chi$L in the LO of $N_c$ as:
\begin{eqnarray}
{\cal L}_{\rm eff}^{\Delta S=0,N_c^2}
&=& 2\bigg(
\tilde{\alpha}_{11}\sum_{i=1}^2V_\mu^i A^{i \mu}
+\tilde{\alpha}_{22}\sum_{i=4}^5V_\mu^i A^{i \mu}\bigg)
+9\tilde{\gamma}_{11}A_\mu^0 V^{0 \mu}
\nonumber\\
&&+3\tilde{\gamma}_{12}
\bigg(-V_\mu^0+2V_\mu^3+\frac{2}{\sqrt{3}}V_\mu^8\bigg)
A^{0 \mu}
+3\tilde{\gamma}_{21}
\bigg(-A_\mu^0+2A_\mu^3+\frac{2}{\sqrt{3}}A_\mu^8\bigg)
V^{0 \mu}
\nonumber\\
&&+\tilde{\gamma}_{22}
\bigg(-V_\mu^0+2V_\mu^3+\frac{2}{\sqrt{3}}V_\mu^8\bigg)
\bigg(-A^{0 \mu}+2A^{3 \mu}+\frac{2}{\sqrt{3}}A^{8 \mu}\bigg).
\label{L-leading}
\end{eqnarray}

\begin{figure}[tbp]
\begin{center}
\includegraphics[height=6cm]{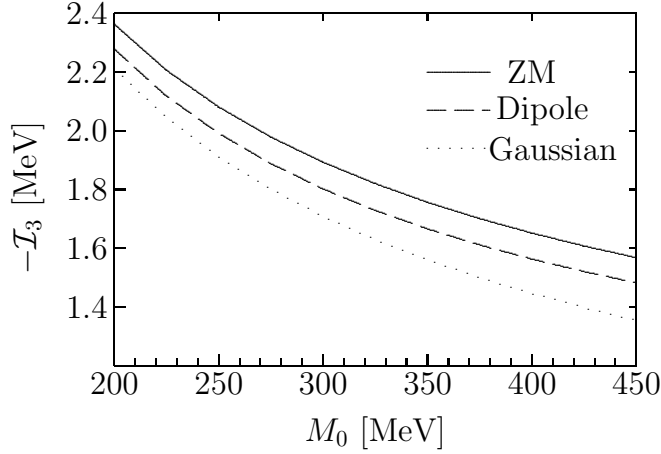}
\caption{The dynamic factor ${\cal I}_3$ as a function of $M_0$.  The solid
curve is drawn with the zero-mode form factor, while the
dashed one depicts the dipole-type form factor.  The dotted one is
for   the Gaussian form factor.}
\label{I3}
\end{center}
\end{figure}

We are now in a position to discuss the LEC in Eq.~(\ref{Leff}), which
consist of the Wilson coefficients, the dynamic factors ${\cal I}_i$,
Cabbibo and Weinberg angles, and the Fermi constant $G_F$, among which
the ${\cal I}_i$ characterize the important feature of the present
approach.  As shown in Eq.~(\ref{eq:coefficient1}), the dynamic factor
${\cal I}_1$ is identified as the quark condensate.
Ref.~\cite{Franz:1999ik} discussed the dependence of the quark
condensate on the $M_0$, where the zero-mode and dipole-type form
factors show similar dependence, while the Gaussian type brings down
the quark condensate noticeably.  The ${\cal I}_2$ is identical to the
Pagels-Stokar pion decay constant given in Eq.(\ref{eq:ps}),
which is approximately $20\, \%$ smaller than the correct $f_\pi^2$
~\cite{Franz:1999yz,Bowler:ir,Golli:1998rf}.
The dynamic factor ${\cal I}_3$ is plotted as a function of $M_0$
in Figure~\ref{I3}.
As in the case of ${\cal I}_1$, the Gaussian-type form factor gives the
smallest value.  It is interesting to compare the present results with
those in the case of the $\Delta S=1$ EW$\chi$L~\cite{Franz:1999ik} 
for which the quark condensate and
${\cal I}_3$-like terms arise from the QCD and electroweak penguin
operators, so that they appear in the LO of the $N_c$ expansion
(${\cal O}(N_c^2)$). However, they are found here in the NLO (${\cal
O}(N_c)$) at the quark level.

Taking into account the current conservation properly, we can express
the LEC in terms of the pion decay constant $f_\pi$, the quark condensate
$\langle \bar{\psi}\psi\rangle$, ${\cal I}_3$, and the Wilson
coefficients:
\begin{eqnarray}
{\cal N}_1 &=& \frac{f_{\pi}^4}{8}    \tilde{\alpha}_{11}, \;\;\;\;
{\cal N}_2  =  \frac{f_{\pi}^4}{8}    \tilde{\alpha}_{22},\;\;\;\;
{\cal N}_3  =  \frac{f_{\pi}^4}{4}    \tilde{\gamma}_{11},\;\;\;\;
{\cal N}_4  =  \frac{f_{\pi}^4}{4}    \tilde{\gamma}_{12},\cr
{\cal N}_{5} &=& \frac{f_{\pi}^4}{4}    \tilde{\gamma}_{21},\;\;\;\;
{\cal N}_{6} =  \frac{f_{\pi}^4}{4} \tilde{\gamma}_{22},\;\;\;\;
{\cal N}_7  =  \frac{f_{\pi}^4}{8N_c}(\tilde{\alpha}_{11}+2\tilde{\beta}_{11}), \;\;\;\;
{\cal N}_8  =  \frac{f_{\pi}^4}{8N_c}(\tilde{\alpha}_{22}+2\tilde{\beta}_{22}),\cr
{\cal N}_9 &=&   4\langle \overline{\psi}\psi\rangle_M {\cal I}_3
(\tilde{\gamma}_{12}+\tilde{\gamma}_{21}+2\tilde{\rho}_{12}+2\tilde{\rho}_{21})
+\frac{f_{\pi}^4}{4 N_c}
(\tilde{\gamma}_{12}-\tilde{\gamma}_{21}+2\tilde{\rho}_{12}-2\tilde{\rho}_{21}),\cr
{\cal N}_{10}  &=&   \frac{f_{\pi}^4}{4 N_c}(\tilde{\gamma}_{22}+\tilde{\rho}_{22})\ .
\label{LECs1}
\end{eqnarray}

The corresponding numerical results are listed in Tables~\ref{tab:lec} and
\ref{tab:lec1}. We find that only
${\cal N}_9$ depends on the type of form factors.
\begin{table}[h]
  \centering
  \begin{tabular}{|c|c|c|c|} \hline
\phantom{xxxx} & $K=1$ & $K=4$ & $K=7$  \\ \hline
${\cal N}_1$ & $7.38$ & $8.31$ & $9.34$  \\
${\cal N}_2$ & $0.39$ & $0.44$ & $0.50$  \\
${\cal N}_3$ & $-0.02$ & $0.51$ & $1.07$ \\
${\cal N}_4$ & $0.01$ & $0.00$ & $0.01$ \\
${\cal N}_{5}$ &$-1.22$ & $-1.40$ & $-1.63$ \\
${\cal N}_{6}$ & $4.13$ & $4.55$ & $5.11$ \\ \hline
${\cal N}_7$ & $2.46$ & $1.26$ & $0.75$  \\
${\cal N}_8$ & $0.13$ & $0.07$ & $0.04$  \\
${\cal N}_9$ & $-7.49$ & $-2.17$ & $0.53$ \\
${\cal N}_{10}$ & $1.38$ & $0.67$ & $0.38$ \\
\hline
\end{tabular}
\caption{Numerical results of the low-energy constants given in unit
of $10^{-5} {\rm MeV}^2$.  The zero-mode form factor is employed with
$M_0=350\,{\rm MeV}$. }
\label{tab:lec}
\end{table}
\begin{table}[h]
\centering
\begin{tabular}{|c|c|c|c|} \hline
Form factor & $K=1$ & $K=4$ & $K=7$  \\ \hline
Dipole & $-6.45$ & $-1.78$ & $0.61$ \\
Gaussian & $-2.81$ & $-0.42$ & $0.90$ \\
\hline
\end{tabular}
\caption{Numerical results for ${\cal N}_9$ are given in unit
of $10^{-5} {\rm MeV}^2$ with the dipole-type and the Gaussian-type form 
factors at $M_0=350\,{\rm MeV}$. ${\cal N}_9$ value for the zero-mode 
form factor is given in Table~\ref{tab:lec}. }
\label{tab:lec1}
\end{table}

There is one last remark: We want to mention that there is a matching
problem between the scale of the effective weak Hamiltonian and that
of the nonlocal chiral quark model from the instanton vacuum.  While
the scale of the effective weak Hamiltonian is determined by the
renormalization point, which is around 1 GeV, that of the nonlocal
chiral quark model comes from the average size of the instanton,
i.e. $1/\rho \simeq 600\,{\rm MeV}$.  Strictly speaking, one has to
match these two different scales~\cite{Bijnens:1997rv}.  However, we
will not consider this problem here, since it is a rather
delicate one and requires a more cautious investigation.

\section{Summary and Conclusions}
In the present work, we concentrated on deriving the $\Delta S = 0$
effective weak Lagrangian incorporating the effective
weak Hamiltonian~\cite{Desplanques:1979hn}.  Based on
the nonlocal chiral quark model from the instanton vacuum, we obtained
the $\Delta S=0$ parity-violating effective weak chiral Lagrangian
with the low-energy constants of the Gasser-Leutwyler type determined.
The dependence of the low-energy constants on the dynamic quark mass
$M_0$ and on the type of form factors was studied.

The effects of the strong interaction were introduced according to the
two different origins: The effect of nonperturbative QCD which is
implemented in the nonlocal chiral quark model from the instanton
vaccum, and the Wilson coefficients which encode the effect of
perturbative gluons~\cite{Desplanques:1979hn}.  In contrast with the
$\Delta S =1$ effective weak chiral
Lagrangian~\cite{Franz:1999yz,Franz:1999ik}, the factorized quark
loops in the integration over the quark field yield
the LO terms, while the unfactorized quark loops do the NLO terms.
We have determined the low-energy constants consisting of the Wilson
coefficients and dynamical quantities such as the pion decay constant
and chiral condensate.  We have estimated the strong enhancement
effects in the LO of the $1/N_c$ expansion.  When it is neglected, our
result turns out to be equivalent with the effective weak chiral
Lagrangian used by Ref.~\cite{Meissner:1998pu}.

The $\Delta S=0$ effective weak chiral Lagrangian in the present work
can be utilized to various strangeness--conserving
weak hadronic processes.  For example, one can derive the weak meson
coupling constants such as $h_\pi^1$. One can also study the parity
violating non--leptonic weak interactions of mesons such as $\eta
\rightarrow \pi^+\pi^-$ or $\eta \rightarrow 2\pi^0$ of which the
upper bound of the decay modes are experimentally known
only~\cite{PDG}.

\section*{Acknowledgments}
CHH and HJL are grateful to B. Desplanques for useful discussions.
HCK thanks M. Musakhanov for valuable discussions.  The work of
HCK and CHH is supported by Korea Research Foundation (Grant
No. KRF-2003-070-C00015).  HJL and CHL are supported by Korea Research
Foundation Grant (KRF-2002-070-C00027) and SEEU-SB2002-0009~(HJL).

\end{document}